\documentclass[aps,pra,showpacs,eqsecnum,twocolumn]{revtex4}
\usepackage{graphicx}
\usepackage{bm}
\begin{document}
\preprint{}
\title{Characterization of quantum angular-momentum fluctuations
via principal components}
\author{\'{A}ngel Rivas and Alfredo Luis}
\email{alluis@fis.ucm.es}
\homepage{http://www.ucm.es/info/gioq}
\affiliation{Departamento de \'{O}ptica, Facultad de Ciencias
F\'{i}sicas, Universidad Complutense, 28040 Madrid, Spain }
\date{\today}
\begin{abstract}
We elaborate an approach to quantum fluctuations of angular
momentum based on the diagonalization of the covariance
matrix in two versions: real symmetric and complex
Hermitian. At difference with previous approaches this is
SU(2) invariant and avoids any difficulty caused by
nontrivial commutators. Meaningful uncertainty relations
are derived which are nontrivial even for vanishing mean
angular momentum. We apply this approach to some relevant
states.
\end{abstract}


\maketitle

\section{Introduction}

Quantum fluctuations and uncertainty relations play
a key role in the fundamentals of quantum physics and
its applications. In this work we focus on angular
momentum variables. Beside mechanics, angular momentum
operators are ubiquitous in areas such as quantum
optics, matter-light interactions, and Bose-Einstein
condensates. Basic observables such as light intensity,
number of particles, and atomic populations are formally
equivalent to angular momentum components
\cite{HL,be,yea,mbe,cvpe}. This is also the case of the
Stokes parameters representing light polarization and
the internal state of two-level atoms identified as
spin 1/2 systems \cite{cvpe,yop,SZ}. Angular momentum
operators are also the generators of basic operations
such as phase shifting, beam splitting, free evolution,
as well as atomic interactions with classical fields
\cite{SZ,yobs}.

Moreover, angular momentum fluctuations are crucial in
diverse areas. This is for example the case of quantum
metrology, implemented by very different physical systems
such as two beam interference with light or Bose-Einstein
condensates, or atomic population spectroscopy. This is
because angular-momentum uncertainty relations determine
the ultimate limit to the resolution of interferometric
and spectroscopic measurements \cite{HL,be,yea}.
A dramatic example has been put forward in Ref. \cite{yea}
concerning atomic clocks based on atomic population
spectroscopy, whose signal to noise ratio is proportional
to the square root of the duration of the measurement.
In such a case, an atomic clock using $10^{10}$ atoms
prepared in an state with reduced angular momentum
fluctuations would yield the same signal to noise ratio
in a measurement lasting 1 second as an atomic clock
with the same number of atoms in an state without reduced
angular momentum fluctuations in a measurement lasting
300 years. From a different perspective it has been shown
that angular momentum fluctuations are useful in the
analysis of many-body entanglement \cite{mbe} and
continuous-variable polarization entanglement \cite{cvpe}.

In comparison to other fundamental variables, such
as position and momentum, the standard uncertainty
relations for angular momentum run into two serious
difficulties: basic commutators are operators instead
of numbers, and there is lack of SU(2) invariance.
Nontrivial commutation relations lead to uncertainty
products bounded by state-dependent quantities.
Among other consequences, these bounds become trivial
for states with vanishing mean angular momentum. On
the other hand, lack of SU(2) invariance is a basic
difficulty since two states connected by a SU(2)
transformation should be equivalent concerning
quantum fluctuations, in the same sense that
phase-space displacements are irrelevant for
position-momentum uncertainty relations.

In this work we elaborate an approach to quantum
fluctuations for angular momentum variables which is
SU(2) invariant and avoids the difficulties caused
by nontrivial commutators. The analysis is based on
the diagonalization of the covariance matrix, that
we study in two versions: real symmetric (Sec. III)
and complex Hermitian (Sec. IV). We examine the
main properties of their eigenvalues (principal
variances) and eigenvectors (principal components).
In Sec. V we derive uncertainty relations
involving principal variances that are meaningful
even in the case of vanishing mean values. Finally,
in Sec. VI we illustrate this approach applying it
to some relevant examples. Section II is devoted
to recall basic concepts and definitions.

\section{Angular momentum operators and SU(2) invariance}

Let us consider arbitrary dimensionless angular
momentum operators $\bm{j} = (j_1, j_2 , j_3 )$
satisfying the commutation relations
\begin{equation}
\label{cr}
[ j_k ,j_\ell ] = i \sum_{n=1}^3 \epsilon_{k,\ell ,n}
j_n ,
\qquad [j_0,\bm{j} ] = \bm{0} ,
\end{equation}
where $\epsilon_{k,\ell,n}$ is the fully antisymmetric
tensor with $\epsilon_{1,2,3} =1$, and $j_0$ is defined
by the relation
\begin{equation}
\label{j2}
\bm{j}^2 = j_0 \left ( j_0 + 1 \right ) .
\end{equation}
For the sake of completeness we take into account that
$j_0$ may be an operator. This is the case of two-mode
bosonic realizations where $j_0$ is proportional to the
number of particles.

More specifically, denoting by $a_{1,2}$ the annihilation
operators of two independent bosonic modes with $[a_j,
a^\dagger_j ]=1$, $[a_1, a_2 ] = [a_1, a^\dagger_2 ]=0$,
we get that the following operators
\begin{eqnarray}
\label{So}
j_0 = \frac{1}{2} \left ( a^\dagger_1 a_1 + a^\dagger_2
a_2 \right ), & &
j_1 = \frac{1}{2} \left ( a^\dagger_2 a_1 + a^\dagger_1
a_2 \right ) , \nonumber \\ & & \\
j_2 = \frac{i}{2} \left ( a^\dagger_2 a_1 - a^\dagger_1
a_2 \right ),  & &
j_3 = \frac{1}{2} \left ( a^\dagger_1 a_1 - a^\dagger_2
a_2 \right ) , \nonumber
\end{eqnarray}
satisfy Eqs. (\ref{cr}) and (\ref{j2}) \cite{SCH}.
Concerning physical realizations, $a_{1,2}$ can represent
the complex amplitude operators of two electromagnetic
field modes. For material systems they can represent the
annihilation operators for two species of atoms in two
different internal states, for example. In any case, the
operator $j_0$ is proportional to the total number of
photons or atoms and $j_3$ is proportional to the number
difference. Therefore, these operators represent basic
detection mechanisms such as the measurement of light
intensity or number of atoms. In this regard we have the
following correspondence
\begin{equation}
|j,m \rangle = |n_1 = j+m \rangle | n_2 = j-m \rangle ,
\end{equation}
between the standard angular momentum basis $|j,m \rangle$
of simultaneous eigenvectors of $j_3$ and $j_0$, i. e.,
$j_3 |j,m \rangle = m |j,m \rangle$, $j_0 |j,m \rangle =
j |j,m \rangle$, and the product of number states in the
two modes $|n_1 \rangle | n_2 \rangle$ with $a^\dagger_j
a_j  |n_j \rangle = n_j | n_j \rangle$.

Angular momentum operators also serve to describe
the internal state of two-level atoms via the definitions
\begin{eqnarray}
j_0 = \frac{1}{2} \left ( |e \rangle \langle e | + |g \rangle
\langle g | \right ), & &
j_1 = \frac{1}{2} \left ( |g \rangle \langle e | + |e \rangle
\langle g | \right ) , \nonumber \\ & & \\
j_2 = \frac{i}{2} \left ( |g \rangle \langle e | - |e \rangle
\langle g | \right ),  & &
j_3 = \frac{1}{2} \left ( |e \rangle \langle e | - |g \rangle
\langle g | \right ) , \nonumber
\end{eqnarray}
where $|e,g \rangle$ are the excited and ground states.
This is formally an spin 1/2 where $j_{0,3}$ represent atomic
populations and $j_{1,2}$ the atomic dipole \cite{SZ}.
Collections of two-level atoms are described by composition
of the individual angular momenta.

Throughout, by SU(2) invariance we mean that the
density operators $\rho$ and $U \rho U^\dagger $ are
fully equivalent concerning quantum fluctuations,
where $U$ is any SU(2) unitary operator exponential
of the angular momentum operators
\begin{equation}
\label{U}
U = \exp \left ( i \theta \bm{u} \cdot \bm{j} \right ),
\end{equation}
with $\theta$ a real parameter, and $\bm{u}$ a unit
three-dimensional real vector. It can be seen that the
action of $U$ on $\bm{j}$ is a rotation $R$ of angle
$\theta$ and axis $\bm{u}$ \cite{CS}
\begin{equation}
\label{rot}
U^\dagger \bm{j} U = R \bm{j} ,
\end{equation}
where $R^t R = R R^t = I$, $I$ is the $3 \times 3$
identity, and the superscript $t$ denotes matrix
transposition. So, the SU(2) invariance is just the
mathematical statement corresponding to the fact
that the conclusions which one could draw from an
angular momentum measurement must be independent of
which set of three orthogonal angular momentum components
one chooses.

One way to guarantee the cited invariance is obtained
by using specific components of the angular momentum
referred to the mean value $\langle\bm{j}\rangle$,
the longitudinal $j_\|$ and transversal $j_{\perp,k}$
components with $k=1,2$. These components are the
projections of $\bm{j}$ on a set of Cartesian axes
adapted to $\langle \bm{j} \rangle$ so that the
longitudinal axis points in the direction of $\langle
\bm{j} \rangle$ \cite{KU93,LK06}. Therefore,
by construction
\begin{equation}
\label{espc}
| \langle \bm{j} \rangle | = \langle j_\| \rangle ,
\qquad
\langle j_{\perp ,k} \rangle = 0  .
\end{equation}
Among other properties $j_\| , j_\perp$ serve to
properly define SU(2) squeezing as reduced
fluctuations of a transversal component
\cite{HL,KU93}. However this approach breaks down
in states with vanishing mean angular momentum.

Finally we recall a basic relation showing that
angular-momentum fluctuations limit the resolution of
interferometric and spectroscopic measurements
\cite{HL,be,yea}. This is because angular momentum
components $j_n$ describe atomic and light free
evolution, as well as the action of linear optical
devices such as beam splitters and interferometers,
via unitary transformations of the form $U_\phi =
\exp ( i \phi j_n )$ acting in a given initial state
$\rho$. The objective of interferometric and
spectroscopic measurements is the accurate determination
of the value of the phase shift $\phi$. This is carried
out by the measurement of a given observable $A$ in the
transformed state $U_\phi \rho U^\dagger_\phi$. In a
simple data analysis, the uncertainty $\Delta \phi$ in
the inferred value of $\phi$ is related to the
fluctuations of the measured observable in the form
\cite{HL}
\begin{equation}
\label{lb}
\Delta \phi = \left |
\frac{\partial \langle A \rangle}{\partial \phi}
\right |^{-1} \Delta A =
\frac{\Delta A }{\left | \langle [A ,j_n ] \rangle
\right |} \geq \frac{1}{2 \Delta j_n },
\end{equation}
where the uncertainty relation $\Delta A \Delta j_n
\geq |\langle [A,j_n ] \rangle |/2$ has been used.
Therefore, the accuracy of the detection is limited
by the fluctuations of the angular momentum component
generating the transformation. In the optimum case
$\Delta j_n \propto \langle j_0 \rangle$ so that
$\Delta \phi$ is inversely proportional to the mean
number of particles. This ultimate limit is known as
Heisenberg limit \cite{HL}. It is worth stressing
that this conclusion applies exclusively to pure
states, as required by the equality in the uncertainty
product between $A$ and $j_n$. These results are
supported by more involved analyses \cite{lb}.

\section{Principal components for real symmetric
covariance matrix}

Complete second-order statistics of the operators
$\bm{j}$ in a given state $\rho$ is contained in
the $3 \times 3$ covariance matrix associated with
$\rho$. Due to the lack of commutativity we can propose
two different covariance matrices: real symmetric
and complex Hermitian.

In this section we focus on the real symmetric covariance
$3 \times 3$ matrix $M$ with  matrix elements
\begin{equation}
\label{Mme}
M_{k, \ell} = \frac{1}{2} \left ( \left \langle
j_k j_\ell \right \rangle + \left \langle j_\ell
j_k \right \rangle \right ) - \left \langle j_k
\right \rangle \left \langle j_\ell  \right \rangle ,
\end{equation}
where mean values are taken with respect to $\rho$ and
we have $M_{k,\ell}^\ast = M_{k,\ell} = M_{\ell ,k}$.
Next we analyze the main properties of $M$.

(i) The covariance matrix $M$ allows us to compute the
variances $( \Delta j_u )^2$ of arbitrary angular-momentum
components $j_u = \bm{u} \cdot \bm{j}$, where $\bm{u}$ is
any unit real vector, in the form
\begin{equation}
\label{utMu}
 \left ( \Delta j_u \right )^2 = \bm{u}^t
M \bm{u}.
\end{equation}

Similarly, $M$ allows us to compute the symmetric
correlation of two arbitrary components $j_u = \bm{u}
\cdot \bm{j}$, $j_v = \bm{v} \cdot \bm{j}$, where
$\bm{u}$, $\bm{v}$ are unit real vectors, in the form
\begin{equation}
\frac{1}{2} \left ( \langle j_u j_v \rangle +
\langle j_v j_u \rangle \right ) -
\langle j_u \rangle \langle j_v \rangle =
\bm{v}^t M \bm{u} = \bm{u}^t M \bm{v}.
\end{equation}

(ii) Since $M$ is real symmetric the transformation
that renders $M$ diagonal is a rotation matrix $R_d$
\begin{equation}
M = R^t_d \pmatrix{ ( \Delta J_1 )^2 & 0 & 0 \cr
 0 & ( \Delta J_2 )^2 & 0 \cr
0 & 0 & ( \Delta J_3 )^2 } R_d .
\end{equation}
The eigenvalues $( \Delta J_k )^2$, $k=1,2,3$, are
the variances of the operators $J_k = \bm{u}_k
\cdot \bm{j}$, where $\bm{u}_k$ are the three real
orthonormal eigenvectors of $M$
\begin{equation}
\label{MD}
M \bm{u}_k =  ( \Delta J_k )^2  \bm{u}_k .
\end{equation}
This implies that $M$ is a positive semidefinite
matrix. Following standard nomenclature in statistics
we refer to $\bm{J}$ and $\Delta \bm{J} = ( \Delta
J_1 ,  \Delta J_2 , \Delta J_3 )$ as principal
components and variances, respectively. We stress
that $\bm{J}$ and $\Delta \bm{J}$ depend on the
system state $\rho$.

(iii) For every $\rho$ the principal components are
uncorrelated
\begin{equation}
\frac{1}{2} \left ( \langle J_k J_\ell \rangle
+ \langle J_\ell J_k \rangle \right ) - \langle
J_k \rangle \langle J_\ell \rangle = 0 ,
\qquad k \neq \ell .
\end{equation}

(iv) The principal components $\bm{J}$ are related
to $\bm{j}$ by the rotation $R_d$ that
diagonalizes $M$
\begin{equation}
\bm{J} = R_d \bm{j} .
\end{equation}
Thus, the three operators $\bm{J}$ are legitimate
mutually orthogonal Hermitian angular-momentum
components satisfying the standard commutation
relations
\begin{equation}
\label{cpc}
[J_k, J_\ell]= i \sum_{n=1}^3 \epsilon_{k,\ell,n} J_n ,
\end{equation}
and
\begin{equation}
\label{Jj}
\bm{J}^2 = \bm{j}^2 , \qquad
\langle \bm{J} \rangle^2 = \langle \bm{j} \rangle^2 .
\end{equation}

(v) At most one principal variance can vanish for
$j_0 \neq 0$ since for $j_0 \neq 0$ no state can be
eigenvector of more than one angular-momentum component.

(vi) The principal variances provide an SU(2) invariant
characterization of fluctuations. The invariance holds
because under any SU(2) transformation (\ref{U}),
(\ref{rot}) we get that $M$ transforms as
$M \rightarrow R M R^t$. Therefore, the covariance matrix
$R M R^t$ associated to the state $U^\dagger \rho U$ has
the same principal variances as the covariance matrix
$M$ associated to the state $\rho$.

(vii) The principal variances are the extremes of the
variances of arbitrary angular momentum components
$j_u = \bm{u} \cdot \bm{j}$ for fixed $\rho$ when the
real unit vector $\bm{u}$ is varied. More specifically,
from Eq. (\ref{utMu}), taking into account that $M$ is
symmetric, and using a Lagrange multiplier $\lambda$
for the constraint $\bm{u}^2 = 1$, we have that the
extremes of $\Delta j_u $ when $\bm{u}$ is varied are
given by the eigenvalue equation
\begin{equation}
M \bm{u}= \lambda \bm{u} ,
\end{equation}
so that from Eq. (\ref{MD}) the extremes coincide
with the principal components.

(viii) Next we examine the relation between the
principal components and the longitudinal $j_\|$
and transversal $j_{\perp,k}$ components, with
$k=1,2$. We can demonstrate that $j_\|$ is a principal
component when the state $\rho$ is invariant
$U \rho U^\dagger = \rho $ under the unitary
transformation $U = \exp (i \pi j_\| )$. This is
because for this transformation
\begin{equation}
U^\dagger j_{\perp,k} U = - j_{\perp,k} ,
\qquad
U^\dagger j_\| U = j_\| ,
\end{equation}
and
\begin{equation}
U^\dagger j_{\perp,k} j_\| U = - j_{\perp,k} j_\| ,
\end{equation}
so that
\begin{equation}
U \rho U^\dagger = \rho \rightarrow \langle j_{\perp,k}
j_\| \rangle = 0 ,
\end{equation}
and similarly for the opposite ordering $\langle j_\|
j_{\perp,k}  \rangle = 0$. Then $j_\|$ is a principal
component and the other two principal components are
transversal.

Invariance under $U = \exp (i \pi j_\| )$ is a very
frequent symmetry. For bosonic two-mode realizations
this is equivalent to symmetry under mode exchange $a_1
\leftrightarrow a_2$ for the modes for which $ j_\|
= (a^\dagger_1 a_2 + a^\dagger_2 a_1 )/2$, where
$a_{1,2}$ are the corresponding annihilation operators.
This is because for these modes $U^\dagger a_1 U
= i a_2$ and $U^\dagger a_2 U = i a_1$ and we have
mode exchange except for a global $\pi/2$ phase change.

(ix) According to Eq. (\ref{lb}) the maximum principal
variance of $\rho$ provides an assessment of the
resolution achievable with $\rho$ in the detection of
small phase shifts generated by angular momentum
components. This includes all linear interferometric
and spectroscopic measurements. Therefore, the maximum
principal variance provides an estimation of the
usefulness of the corresponding state in quantum
metrology.

(x) Next we derive upper and lower bounds for the
principal variances of states with $\langle \bm{j}
\rangle = \bm{0}$. In such a case from Eqs. (\ref{j2})
and (\ref{Jj}) we have
\begin{equation}
\label{tr}
\textrm{tr} M = \left ( \Delta J_1 \right )^2 +
\left ( \Delta J_2 \right )^2 + \left ( \Delta J_3
\right )^2 = \langle j_0 \left ( j_0 + 1 \right )
\rangle .
\end{equation}
If we arrange the principal variances in decreasing
order $\Delta J_1 \geq \Delta J_2 \geq \Delta J_3$,
and using Eq. (\ref{tr}) we get the following bounds
for the principal variances
\begin{eqnarray}
\label{do}
& \langle j^2_0 \rangle \geq \left ( \Delta J_1
\right )^2 \geq \frac{1}{3} \left \langle j_0 \left (
j_0 + 1 \right ) \right \rangle ,
& \nonumber \\
& \frac{1}{2} \left \langle j_0 \left ( j_0 + 1 \right )
\right \rangle \geq \left ( \Delta J_2 \right )^2 \geq
\frac{1}{2} \langle j_0 \rangle ,
& \nonumber \\
& \frac{1}{3} \left \langle j_0 \left ( j_0 + 1 \right )
\right \rangle \geq \left ( \Delta J_3 \right )^2
\geq 0 . &
\end{eqnarray}
The upper bound for $\Delta J_1 $ holds because for
an arbitrary component $\langle j_k^2 \rangle \leq
\langle j_0^2 \rangle$, while the lower bound is reached
when all the principal variances are equal. The upper
bound for $\Delta J_2 $ is reached when $\Delta J_1 =
\Delta J_2$ and $\Delta J_3 =0$, while the lower bound
is reached when $(\Delta J_1 )^2 = \langle j_0^2 \rangle $
and $\Delta J_2 = \Delta J_3$. Finally, the upper bound
for $\Delta J_3 $ is reached when all the principal
variances are equal, while the lower bound occurs for
$\Delta J_3 = 0$.

(xi) The lower bound for $\Delta J_1 $ in Eq. (\ref{do})
is, roughly speaking, of the order of $\langle j_0 \rangle$
so that from Eq. (\ref{lb}) all pure states with $\langle
\bm{j} \rangle = \bm{0}$ can reach maximum interferometric
precision (Heisenberg limit). This explains why most
optimum states for metrological applications satisfy
$\langle \bm{j} \rangle = \bm{0}$ (see Sec. VI).

\section{Principal components for complex Hermitian covariance matrix}

In this section we elaborate the statistical description
of angular-momentum fluctuations via the complex
Hermitian $3 \times 3$ covariance matrix $\tilde{M}$ with
matrix elements
\begin{equation}
\tilde{M}_{k, \ell} = \left \langle j_k j_\ell \right
\rangle - \left \langle j_k \right \rangle \left
\langle j_\ell  \right \rangle ,
\end{equation}
with $\tilde{M}_{k,\ell}^\ast = \tilde{M}_{\ell ,k}$.
The two matrices $M$ and $\tilde{M}$ contain essentially
the same information since they only differ by a factor
$\langle \bm{j} \rangle$
\begin{equation}
\label{rel}
M_{k,\ell} = \tilde{M}_{k,\ell} - \frac{i}{2}
\sum_{n=1}^3 \epsilon_{k,\ell,n} \langle j_n \rangle .
\end{equation}
In particular they coincide exactly for all states with
$\langle \bm{j} \rangle = \bm{0}$.

Nevertheless, $M$ and  $\tilde{M}$ are different
matrices so we can exploit this difference by
examining the most relevant features that they do
not share in common.

(i) The covariance matrix $\tilde{M}$ provides the
properly defined variances of complex linear
combinations of angular-momentum components $j_u =
\bm{u} \cdot \bm{j}$ for arbitrary unit complex vectors
$\bm{u}$ with $\bm{u}^\dagger \bm{u} = 1$.

More specifically, for $\bm{u}^\ast \neq \bm{u}$ we
have that $j_u$ is not Hermitian, $j_u^\dagger \neq
j_u$,  so that the variance must be redefined, for
example in the form \cite{LL76}
\begin{equation}
\label{cv}
\left ( \Delta j_u \right )^2 = \langle j_u^\dagger j_u
\rangle - | \langle j_u \rangle |^2 .
\end{equation}
Then $\tilde{M}$ provides $\Delta j_u $ as
\begin{equation}
\label{tp}
\left ( \Delta j_u \right )^2 = \bm{u}^\dagger
\tilde{M} \bm{u} .
\end{equation}
Note that for real $\bm{u}$ we have
\begin{equation}
\label{rc}
\bm{u}^\dagger \tilde{M} \bm{u} = \bm{u}^t M \bm{u} .
\end{equation}

Similarly, we can compute the correlation of two complex
projections $j_u = \bm{u} \cdot \bm{j}$, $j_v = \bm{v}
\cdot \bm{j}$ where $\bm{u}$, $\bm{v}$ are arbitrary
unit complex vectors, in the form
\begin{equation}
\label{cpH}
\langle j^\dagger_v j_u \rangle - \langle j^\dagger_v
\rangle \langle j_u \rangle = \bm{v}^\dagger
\tilde{M} \bm{u} .
\end{equation}

(ii) Since $\tilde{M}$ is Hermitian it becomes diagonal
by means of a $3 \times 3$ unitary matrix
$\mathcal{U}_d$
\begin{equation}
\tilde{M} = \mathcal{U}^\dagger_d \pmatrix{
( \Delta \tilde{J}_1 )^2 & 0 & 0 \cr
 0 & ( \Delta \tilde{J}_2 )^2 & 0 \cr
0 & 0 & ( \Delta \tilde{J}_3 )^2 } \mathcal{U}_d ,
\end{equation}
where the elements on the diagonal are the variances
(\ref{cv}) of the operators $\tilde{J}_k = \bm{u}_k
\cdot \bm{j}$, where $\bm{u}_k$ are the three complex
orthonormal eigenvectors of $\tilde{M}$
\begin{equation}
\label{tMD}
\tilde{M} \bm{u}_k =  ( \Delta \tilde{J}_k )^2
\bm{u}_k .
\end{equation}
This implies that $\tilde{M}$ is positive semidefinite.
We again refer to $\tilde{\bm{J}}$ and $\Delta
\tilde{\bm{J}} = ( \Delta \tilde{J}_1 ,  \Delta
\tilde{J}_2 , \Delta \tilde{J}_3 ) $ as principal
components and variances, respectively. Since a global
phase is irrelevant we regard always $\mathcal{U}_d$
as an SU(3) matrix.

(iii) For every state $\rho$ the principal components are
uncorrelated
\begin{equation}
\langle \tilde{J}^\dagger_k \tilde{J}_\ell \rangle -
\langle \tilde{J}^\dagger_k \rangle \langle \tilde{J}_\ell
\rangle = 0,
\qquad k \neq \ell .
\end{equation}

(iv) Standard commutation relations (\ref{cr}) are not
preserved under transformations by unitary $3 \times 3$
matrices
\begin{equation}
\label{JUj}
\tilde{\bm{J}} = \mathcal{U} \bm{j} .
\end{equation}
However, a slight modification of (\ref{cr}) is actually
preserved under the action (\ref{JUj}) of SU(3) matrices.
These are
\begin{equation}
\label{nscr}
[\tilde{J}_k ,\tilde{J}_\ell ] = i
\sum_{n=1}^3 \epsilon_{k,\ell ,n} \tilde{J}_n^\dagger  ,
\qquad
[j_0,\bm{\tilde{J}} ] = \bm{0} ,
\end{equation}
which are equivalent to Eq. (\ref{cr}) for Hermitian
operators. We have also
\begin{equation}
\label{tJj}
\tilde{\bm{J}}^\dagger \cdot \tilde{\bm{J}} = \bm{j}^2 ,
\qquad \langle \tilde{\bm{J}} \rangle^\ast \cdot
\langle \tilde{\bm{J}} \rangle = \langle \bm{j}
\rangle^2 .
\end{equation}
The preservation of (\ref{nscr}) under the action of SU(3)
matrices in Eq. (\ref{JUj}) can be demonstrated by direct
computation of the commutators after expressing the most
general SU(3) matrix $\mathcal{U}$ as a suitable product
of matrices belonging to the SU(2) subgroup of the form
\cite{SU3SU2}
\begin{equation}
\label{form}
\mathcal{U} = \pmatrix{ \cos \theta e^{i \phi} &
\sin \theta e^{i \varphi} & 0 \cr - \sin \theta
e^{-i \varphi} & \cos \theta e^{-i \phi}& 0
\cr 0 & 0 & 1} ,
\end{equation}
and permutations of lines and columns. Each matrix
(\ref{form}) transforms operators $\tilde{\bm{J}}$
satisfying Eq. (\ref{nscr}) into another set of
operators $\tilde{\bm{J}}^\prime = \mathcal{U}
\tilde{\bm{J}}$ fulfilling the same commutation
relations (\ref{nscr}).

(v) From Eq. (\ref{rel}) we can derive the equality
of traces $\textrm{tr} M = \textrm{tr} \tilde{M}$
so that
\begin{equation}
\label{sfe}
\left ( \Delta \tilde{\bm{J}} \right )^2 =
\left ( \Delta \bm{J} \right )^2 =
\left ( \Delta \bm{j} \right )^2 =
\langle j_0 \left ( j_0 + 1 \right ) \rangle
- \langle \bm{J} \rangle^2 ,
\end{equation}
with
\begin{equation}
\label{sfe2}
\left | \langle \tilde{\bm{J}} \rangle \right |^2
= \langle \bm{J} \rangle^2 = \langle \bm{j}
\rangle^2 .
\end{equation}
Since $\langle \bm{J} \rangle^2 \leq \langle j_0
\rangle^2$ we have $( \Delta \tilde{\bm{J}} )^2 >0$.
This implies that there must be at least a
nonvanishing principal variance. Otherwise, two
principal variances of $\tilde{M}$ can vanish
simultaneously for the same state, as shown in
Sec. VI for the SU(2) coherent states.

(vi) Although the traces of $M$ and $\tilde{M}$
are equal the determinants are different. This can
be easily proven by expressing $\tilde{M}$ in the
principal-component basis that renders $M$ diagonal
\begin{equation}
\tilde{M} = \pmatrix{
( \Delta J_1 )^2 & i \langle J_3 \rangle /2 & - i
\langle J_2 \rangle /2 \cr
- i \langle J_3 \rangle /2 & ( \Delta J_2 )^2 &
i \langle J_1 \rangle /2 \cr
i \langle J_2 \rangle /2 & - i \langle J_1 \rangle /2
& ( \Delta J_3 )^2 } ,
\end{equation}
so that
\begin{equation}
\textrm{det} \tilde{M} = \textrm{det} {M} - \frac{1}{4}
\sum_{k=1}^3 ( \Delta J_k )^2 \langle J_k \rangle ^2,
\end{equation}
and $\textrm{det}M \geq \textrm{det} \tilde{M}$.

(vii) The principal variances of $\tilde{M}$ are SU(2)
invariant since under SU(2) transformations we have
$\tilde{M} \rightarrow R \tilde{M} R^t$, so that the
covariance matrix $R \tilde{M} R^t$ for the state
$U^\dagger \rho U$ has the same eigenvalues as the
covariance matrix $\tilde{M}$ for the state $\rho$.

(viii) The principal variances are the extremes of the
variances $(\Delta j_u )^2$ of any complex combination
of angular momentum components $j_u = \bm{u} \cdot
\bm{j}$ where $\bm{u}$ is a complex unit vector. More
specifically, from Eq. (\ref{tp}), and introducing a
Lagrange multiplier $\lambda$ to take into account the
constraint $\bm{u}^\dagger \cdot \bm{u}= 1$, we get
that the extremes of $\Delta j_u $ are given by the
eigenvalue equation
\begin{equation}
\tilde{M} \bm{u}= \lambda \bm{u} ,
\end{equation}
so that from Eq. (\ref{tMD}) the extremes of
$(\Delta j_u )^2$ are the principal variances
$\Delta \tilde{\bm{J}}$.

(ix) The principal variances of $\tilde{M}$ are more
extreme than the principal variances of $M$ since
from Eq. (\ref{rc}) the variation process for the
complex Hermitian case takes place over a larger
set of operators $j_u$, with complex $\bm{u}$, that
includes as a particular case the projections on
real $\bm{u}$.

(x) Because of Eqs. (\ref{sfe}) and (\ref{sfe2}) for
$\langle \bm{j} \rangle = \bm{0}$ the upper and
lower bounds for principal variances in Eq. (\ref{do})
also hold replacing $\bm{J}$ by $\tilde{\bm{J}}$.

(xi) It is questionable whether $\Delta j_u$ for
$j_u^\dagger \neq j_u$ represents practical observable
fluctuations. For example, for $j_u = j_1 + i j_2$
we have
\begin{equation}
\left ( \Delta j_u \right )^2 = \left ( \Delta
j_1 \right )^2 + \left ( \Delta j_2 \right )^2
- \langle j_3 \rangle ,
\end{equation}
so we can have $\Delta j_u =0$ with $\Delta j_{1,2}
\neq 0$, being this the case of the SU(2) coherent
states (see Sec. VI). Nevertheless, from
Eq. (\ref{sfe}) we have that $\Delta \tilde{\bm{J}}$
contains all angular momentum fluctuations.

In this regard it is worth recalling that non Hermitian
operators can be related to experimental processes, as
demonstrated by double homodyne detection where the
statistics is given by projection on quadrature coherent
states \cite{dh}. In our context, the eigenstates of
$j_u = j_1 + i j_2$ are SU(2) coherent states that
define by projection the SU(2) $Q$ function \cite{CS}.
This is an observable probability distribution function,
via double homodyne detection of two field modes for
example \cite{mQ}.

\section{Uncertainty relations}

Variances are the most popular building blocks of
uncertainty relations. For angular momentum, the
standard procedure leads to
\begin{equation}
\label{sup}
\Delta j_1 \Delta j_2 \geq \frac{1}{2} | \langle j_3
\rangle | ,
\end{equation}
and cyclic permutations. This uncertainty relation
faces two difficulties. On the one hand it is bounded by
an state-dependent quantity that vanish for states with
$\langle \bm{j} \rangle =\bm{0}$. On the other hand, it
lacks SU(2) invariance so that it leads to different
conclusions when applied to SU(2) equivalent states
\cite{WE85}. These difficulties can be avoided by using
the principal variances leading to meaningful SU(2)
invariant relations which are nontrivial even for states
with $\langle \bm{j} \rangle = \bm{0}$.

A first SU(2) invariant uncertainty relation can be
derived from the trace of $M$ (or equivalently
$\tilde{M}$) \cite{DE77}
\begin{equation}
\label{Ds}
\textrm{tr} M = \left ( \Delta \bm{J} \right )^2 =
\left \langle j_0 \left ( j_0 + 1 \right ) \right
\rangle - \langle \bm{J} \rangle^2 \geq \langle j_0
\rangle ,
\end{equation}
where we have used that for any component $ \langle j_k
\rangle^2 \leq \langle j_k^2 \rangle \leq \langle j_0^2
\rangle$.

The minimum uncertainty states with $\left ( \Delta
\bm{J} \right )^2 = \langle j_0 \rangle$ are obtained
for maximum $\langle \bm{J} \rangle^2 = \langle j^2_0
\rangle$. This is satisfied exclusively by SU(2)
coherent states \cite{CS,DE77}
\begin{equation}
| j,  \theta, \bm{u} \rangle = U (\theta, \bm{u})
| j, j \rangle ,
\end{equation}
where $U$ is any SU(2) unitary operator (\ref{U}) and
$| j, m \rangle$ are the simultaneous eigenvectors of
$j_0$ and $j_3$, with eigenvalues $j$ and $m$,
respectively. On the other hand, maximum uncertainty
$\left ( \Delta \bm{J} \right )^2$ is obtained for
$\langle \bm{J} \rangle = \bm{0}$. This is the case
of the state $|j,m = 0 \rangle$ for example (see Sec. VI).

The uncertainty relation (\ref{Ds}) is nontrivial even
for $ \langle \bm{j} \rangle = \bm{0}$. Nevertheless,
this is not very informative about angular momentum
statistics since this is actually just a function of
the first moments $\langle \bm{j} \rangle$. In order
to proceed further deriving more meaningful uncertainty
relations let us split the analysis in two cases
$\langle \bm{j} \rangle \neq \bm{0}$ and $\langle \bm{j}
\rangle = \bm{0}$.

\subsection{Case $\langle \bm{j} \rangle \neq \bm{0}$}

\subsubsection{Product of variances}

For $\langle \bm{j} \rangle \neq \bm{0}$ an SU(2)
invariant product of variances can be derived by
applying the standard procedure to the longitudinal
and transversal components (\ref{espc}), leading to
just one nontrivial relation \cite{LK06}
\begin{equation}
\label{ur}
\Delta j_{\perp,1} \Delta j_{\perp,2} \geq
\frac{1}{2} | \langle j_\| \rangle | ,
\end{equation}
while the other two are trivial $\Delta j_{\perp,k}
\Delta j_{\|} \geq 0$, for $k=1,2$.

When $J_\| = j_\|$ is a principal component we can show
that the principal transversal variances $\Delta
J_{\perp ,k}$ provide the minimum uncertainty product
\begin{equation}
\label{ptc}
\Delta j_{\perp,1} \Delta j_{\perp,2} \geq
\Delta J_{\perp ,1} \Delta J_{\perp ,2}
\geq \frac{1}{2} | \langle j_\| \rangle |.
\end{equation}
This is because the determinant of $M$ is invariant
under rotations of $\bm{j}$, and the principal
components are uncorrelated. Moreover they are the
extreme variances in the transversal plane according
to point (vii) in Sec. III.

\subsubsection{Sum of variances}

We can begin with by particularizing Eq. (\ref{j2})
to longitudinal and transversal components leading to
\begin{equation}
\left ( \Delta j_{\perp ,1} \right )^2 + \left (
\Delta j_{\perp ,2} \right )^2 = \langle j_0
\left ( j_0 + 1 \right ) \rangle - \langle j_{\|}^2
\rangle .
\end{equation}
Since we have always $\langle j^2_0 \rangle \geq
\langle j_{\|}^2 \rangle $ we get the following
lower bound to the sum of transversal variances
\begin{equation}
\label{sv}
\left ( \Delta j_{\perp ,1} \right )^2 + \left (
\Delta j_{\perp ,2} \right )^2 \geq \langle j_0
\rangle ,
\end{equation}
where the equality is reached by the SU(2) coherent
states exclusively.

Let us note that this relation is stronger than the
similar one that can be derived from Eq. (\ref{ur}),
\begin{equation}
\left ( \Delta j_{\perp ,1} \right )^2 + \left (
\Delta j_{\perp ,2} \right )^2 \geq | \langle j_\|
\rangle | ,
\end{equation}
since $ \langle j_0 \rangle \geq | \langle j_\|
\rangle |$ always.

Equation (\ref{sv}) holds irrespectively of whether
$j_\| $ is a principal component or not. Furthermore,
when $j_\|$ is a principal component of $M$ or
$\tilde{M}$ we have, respectively
\begin{eqnarray}
\label{sv2}
& \left ( \Delta J_{\perp ,1} \right )^2 + \left (
\Delta J_{\perp ,2} \right )^2 \geq \langle j_0
\rangle , & \nonumber \\
& \left ( \Delta \tilde{J}_{\perp ,1} \right )^2 +
\left ( \Delta \tilde{J}_{\perp ,2} \right )^2 \geq
\langle j_0 \rangle . &
\end{eqnarray}

\subsection{Case $\langle \bm{j} \rangle = \bm{0}$}

States with $\langle \bm{j} \rangle = \bm{0}$ arise
very often in quantum metrological applications
as explained in point (xi) of Sec. III
\cite{HL,nn,noon}. In such a case the standard
uncertainty products (\ref{sup}) are all trivial
$\Delta j_k  \Delta j_\ell  \geq 0$ since they
do not establish any lower bound to the product
of variances. Moreover, the components $j_\| $,
$j_\perp$ are undefined.

\subsubsection{Product of variances}

We can derive a suitable lower bound for the product
$\Delta J_1 \Delta J_2 $ of the two larger principal
variances of $M$ (or $\tilde{M}$) with $\Delta J_1
\geq \Delta J_2 \geq \Delta J_3 $, valid for all states
with $\langle \bm{j} \rangle = \bm{0}$.

To this end we begin with by considering the minimum
of $\Delta J_2$ for fixed $\Delta J_1$. From the
equality (\ref{tr}) we get
\begin{equation}
\left ( \Delta J_2 \right )^2 + \left ( \Delta J_3
\right )^2 =  \left \langle j_0 \left ( j_0 + 1
\right ) \right \rangle - \left ( \Delta J_1
\right )^2 ,
\end{equation}
and for fixed $\Delta J_1$ the sum $( \Delta J_2 )^2
+ ( \Delta J_3 )^2$ is constant. Taking into account
that $\Delta J_2 \geq \Delta J_3$ we get that the
minimum $\Delta J_2$ occurs when
\begin{equation}
\Delta J_2 = \Delta J_3 =  \frac{1}{2} \left [
\left \langle j_0 \left ( j_0 + 1 \right )
\right \rangle - \left ( \Delta J_1 \right )^2
\right ] ,
\end{equation}
and then it holds that
\begin{equation}
\label{prim}
\left ( \Delta J_2 \right )^2 \left ( \Delta J_1
\right )^2 \geq \frac{1}{2} \left ( \Delta J_1
\right )^2 \left [ \left \langle j_0 \left ( j_0
+ 1 \right ) \right \rangle - \left ( \Delta J_1
\right )^2 \right ] .
\end{equation}

The minimum of the right-hand side takes place
when $\Delta J_1 $ reaches its extremes values
in Eq. (\ref{do}) so that
\begin{equation}
\left ( \Delta J_2 \right )^2 \left ( \Delta J_1
\right )^2 \geq \textrm{min} \left [ \frac{1}{2}
\langle j_0 \rangle \langle j^2_0 \rangle ,
\frac{1}{9} \langle j_0 (j_0 + 1 ) \rangle^2
\right ] ,
\end{equation}
where ``min'' refers to the minimum of the
alternatives.

For $\langle j_0 \rangle \geq 2$ we get that this is
always
\begin{equation}
\label{prim2}
\left ( \Delta J_2 \right )^2 \left ( \Delta J_1
\right )^2 \geq \frac{1}{2} \left \langle j_0 \right
\rangle \left \langle j^2_0 \right \rangle ,
\end{equation}
and the equality is reached for states with maximum
$\Delta J_1$ and minimum $\Delta J_2$
\begin{equation}
\label{cm}
\left ( \Delta J_1 \right )^2 = \left \langle j^2_0
\right \rangle, \quad
\left ( \Delta J_2 \right )^2 =
\left ( \Delta J_3 \right )^2 = \frac{1}{2} \left
\langle j_0 \right \rangle .
\end{equation}
We will see in the next section that this is the case
of the Schr\"{o}dinger cat states (\ref{scs}).

\subsubsection{Sum of variances}

Nontrivial bounds to sums of variances can be derived
by particularizing Eq. (\ref{j2}) to three components
$j_u$, $j_v$, $j_w$ obtained by projection on three
mutually orthogonal (complex in general) unit
vectors $\bm{u}$, $\bm{v}$, $\bm{w}$
\begin{equation}
\label{Ds2}
\left ( \Delta j_u \right )^2 + \left ( \Delta j_v
\right )^2 + \left ( \Delta j_w \right )^2 =
\left \langle j_0 \left ( j_0 + 1 \right )
\right \rangle .
\end{equation}
Since for any projection, say $j_w$, we have $\langle
j_0^2 \rangle \geq \langle j^\dagger_w j_w \rangle =
(\Delta j_w )^2$ we get
\begin{equation}
\label{fap}
\left ( \Delta j_u \right )^2 + \left ( \Delta j_v
\right )^2 \geq \langle j_0 \rangle ,
\end{equation}
where $\bm{u}$, $\bm{v}$ are orthogonal complex
unit vectors $\bm{u}^\dagger \cdot \bm{v} =0$.
As a byproduct we obtain that for $\langle \bm{j}
\rangle = \bm{0}$ only one principal variance
$\Delta \tilde{J}$ can vanish.

Finally we can appreciate that the above relations
involve the trace of the covariance matrix being
derived seemingly without resorting to commutation
relations. Nevertheless, commutation relations
are also at the hearth of these uncertainty relations
since this is the ultimate reason forbidding the
simultaneous vanishing of the fluctuations of all
angular momentum components. In this regard, as shown
in the original Schr\"{o}dinger's paper
position-momentum uncertainty relations can be
fruitfully related to the corresponding covariance
matrix \cite{Sch}.

\section{Examples}

In this section we apply the preceding formalism to
some relevant and illustrative quantum states.

\subsection{States $|j,m \rangle$}

Let us consider the simultaneous eigenstates $|j,m
\rangle$ of $j_0$ and $j_3$ with eigenvalues $j$
and $m$ respectively. This family includes the SU(2)
coherent states for $m= \pm j$ and the limit of
SU(2) squeezed coherent states for $m=0$ \cite{HL,CS}.
For two-mode bosonic realizations the case  $m=0$ is
the product of states with the same definite number
of particles in each mode \cite{nn}.

For $m \neq 0$ we have $\langle \bm{j} \rangle \neq
\bm{0}$ and there is a longitudinal component with
$j_\| = j_3$. In such a case $U \rho U^\dagger =
\rho$ for $U = \exp ( i \pi j_\| )$ and $j_\|$ is
a principal component.

The real symmetric covariance matrix is directly
diagonal in any basis containing the longitudinal
component $j_3 = j_\|$
\begin{equation}
\label{mjm}
M = \frac{1}{2} \pmatrix{ j(j+1) - m^2 & 0 & 0 \cr
0 &  j(j+1) - m^2 & 0 \cr 0 & 0 & 0} ,
\end{equation}
with principal variances
\begin{equation}
\label{pvsc}
( \Delta J )^2 = 0, \frac{1}{2} \left [ j(j+1) -
m^2 \right ] .
\end{equation}

On the other hand, the complex Hermitian covariance
matrix is
\begin{equation}
\label{tmjm}
\tilde{M} = \frac{1}{2} \pmatrix{ j(j+1) - m^2 & i m
& 0 \cr - im &  j(j+1) - m^2 & 0 \cr 0 & 0 & 0} ,
\end{equation}
with principal components
\begin{eqnarray}
\label{opc}
& \tilde{J}_{\perp, 1} = \frac{1}{\sqrt{2}} j_+ =
\frac{1}{\sqrt{2}} \left ( j_1 + i j_2 \right ), &
\nonumber \\
& \tilde{J}_{\perp ,2} = \frac{1}{\sqrt{2}} j_-  =
\frac{1}{\sqrt{2}} \left ( j_1 - i j_2 \right ), &
\nonumber \\ & \tilde{J}_\| = j_3 , &
\end{eqnarray}
so that $\tilde{J}_{\perp ,1,2}$ are proportional to the
ladder operators  $j_\pm$. The principal variances are
\begin{eqnarray}
\label{pvhc}
& ( \Delta \tilde{J}_{\perp, 1} )^2 = \frac{1}{2}
\left [ j(j+1) - m (m + 1) \right ] , & \nonumber \\
& ( \Delta \tilde{J}_{\perp ,2} )^2 = \frac{1}{2}
\left [ j(j+1) - m (m - 1) \right ] , & \nonumber \\
& ( \Delta \tilde{J}_\| )^2 =  0 , &
\end{eqnarray}
which are larger and lesser, respectively, than
the variances (\ref{pvsc}) of the real symmetric
case, in accordance with point (ix) of Sec. IV.

The SU(2) coherent states are minimum uncertainty
states for the sum of three variances in
Eq. (\ref{Ds}), and for the product and sum of
variances of transversal components in
Eqs. (\ref{ptc}), (\ref{sv}), and (\ref{sv2}).
The scaling of the largest principal variance as
$(\Delta J_1 )^2 \propto j$ agrees with the fact
that the SU(2) coherent states are not optimum for
metrological applications. Optimum states scaling
as $(\Delta J_1 )^2 \propto j^2$ can be found below
in this section.

For SU(2) coherent states two of the principal
variances in Eq. (\ref{pvhc}) vanish. This
corresponds to the  fact that they satisfy the
double eigenvalue relation $j_\pm |j, \pm j
\rangle = 0$ and $j_3 |j, \pm j \rangle = \pm j
|j, \pm j \rangle$, so that $|j, \pm j \rangle$
are eigenstates of two of the principal components
in Eq. (\ref{opc}).

On the other hand, for the states with $m \neq
\pm j$ the nonvanishing variances increase for
decreasing $|m|$ and for $m=0$ the maximum scales
as $(\Delta J_1 )^2 \propto j^2$, which is consistent
with the usefulness of these states in quantum
metrology, in agreement with points (ix) and (xi)
in Sec. III \cite{nn}. Moreover, the states $m=0$
are far from the lower bounds of the uncertainty
relations in Eqs. (\ref{Ds}), (\ref{prim2}), and
(\ref{fap}).

\subsection{SU(2) squeezed coherent states}

Let us consider the SU(2) squeezed coherent states
defined by the eigenvalue equations \cite{HL,KU93}
\begin{equation}
\label{ee}
\left ( j_{\perp, 1} + i \xi j_{\perp, 2 }\right )
| \xi \rangle = 0 ,
\qquad
j_0 | \xi \rangle = j | \xi \rangle ,
\end{equation}
where $\xi$ is a real parameter with $\xi \geq 0$
without loss of generality. The first of these
equations corresponds to the case of zero eigenvalue
among a larger family of eigenvalue equations
\cite{HL}. The states $| \xi \rangle$ are fully
defined by the eigenvalue relations (\ref{ee}).
An approximate solution is provided below in Eq. (\ref{sc}).
For $\xi =1$ these states are the SU(2) coherent states
while for $\xi \neq 1$ they are SU(2) squeezed coherent
states being minimum uncertainty states of the uncertainty
product (\ref{ptc}) with reduced fluctuations in the
component $j_{\perp, 1}$ for $\xi <1$. They satisfy the
squeezing criterion suitable for interferometric and
spectroscopic measurements approaching the Heisenberg
limit \cite{HL,KU93}.

It is worth stressing that for any $\xi$ the vanishing
of the eigenvalue in Eq. (\ref{ee}) grants that
$j_{\perp,k}$, $k=1,2$, are actually transversal
components for all parameters $\xi$. This can be
readily seen by projecting Eq. (\ref{ee}) on
$| \xi \rangle$.

Furthermore, we can show that Eq. (\ref{ee})
grants also that the operators $j_{\perp,k}$ in
Eq. (\ref{ee}) and $j_\| = -i [ j_{\perp,1} ,
j_{\perp, 2}]$ are principal components. To this
end we can project Eq. (\ref{ee}) on $j_\| | \xi
\rangle$ leading to
\begin{equation}
\label{pro}
\langle \xi | j_\| j_{\perp, 1} | \xi \rangle
+ i \xi \langle \xi | j_\| j_{\perp, 2} | \xi
\rangle  = 0 .
\end{equation}
The commutation relations and $\langle j_{\perp, k}
\rangle =0$ imply that  $\langle j_\| j_{\perp,k}
\rangle = \langle j_{\perp,k} j_\| \rangle$ so that
the mean values in Eq. (\ref{pro}) are real
quantities. Thus Eq. (\ref{pro}) implies that both
mean values vanish and $j_\|$ is a principal component
both for $M$ and $\tilde{M}$. Moreover, by adding
the projections on $j_{\perp,1} | \xi \rangle$
and $- i \xi j_{\perp,2} | \xi \rangle$ we get that
$j_{\perp,k}$ are uncorrelated in the sense that
$\langle \xi | ( j_{\perp,1} j_{\perp,2} +
j_{\perp,2} j_{\perp, 1} )| \xi \rangle = 0$.

Therefore, $M$ is diagonal in the $j_{\perp,k},
j_\| $ basis so they are the principal components
of $M$. The vanishing of the eigenvalue in
Eq. (\ref{ee}) is the only possibility dealing
with principal components since otherwise the
correlations between components are proportional to
the eigenvalue, spoiling property (iii) in Sec. III
\cite{DND05}.

All this suggests that Eq. (\ref{ee}) may be taken
as the proper SU(2) invariant form of defining the
SU(2) squeezed coherent states. The exact solution
of Eq. (\ref{ee}) for arbitrary $\xi$ is difficult
to handle \cite{HL}. For definiteness we can consider
the limit $\xi \rightarrow 0$ retaining the first
nonvanishing power on $\xi$. In the basis $|j,m \rangle$
of eigenvectors of $j_0$ and $j_{\perp, 1}$ we have
\begin{equation}
\label{sc}
| \xi \rangle \simeq N \left [ |j,0 \rangle -
\frac{i}{2} \xi \sqrt{j(j+1)} \left ( | j, 1
\rangle - | j, -1 \rangle \right ) \right ] ,
\end{equation}
where $N$ is a normalization constant.

In this approximation, the principal variances of $M$
are
\begin{eqnarray}
& ( \Delta J_{\perp,1} )^2 \simeq \frac{1}{2}j(j+1)
\xi^2 , & \nonumber \\
& ( \Delta J_{\perp,2} )^2 \simeq ( \Delta J_\| )^2
\simeq \frac{1}{2}j(j+1), &
\end{eqnarray}
with
\begin{equation}
\langle J_\| \rangle \simeq  j(j+1) \xi .
\end{equation}

This is a minimum uncertainty state for the uncertainty
product in Eq. (\ref{ptc}) while for the sums of
variances it behaves essentially as the state $|j, m=0
\rangle$. The metrological usefulness of these states
is confirmed by the scaling of the maximum principal
variance as $(\Delta J_1 )^2 \propto j^2$, in accordance
with point (ix) in Sec. III.

The complex Hermitian covariance matrix is no longer
diagonal in the $j_{\perp,k}$ basis
\begin{equation}
\tilde{M} =  \frac{j(j+1)}{2}
\pmatrix{ \xi^2 & i \xi & 0 \cr
- i \xi & 1 & 0 \cr 0 &  0 & 1} .
\end{equation}
The principal components are
\begin{eqnarray}
& \tilde{J}_{\perp ,1} = j_{\perp,1}  +
i \xi j_{\perp,2}, & \nonumber \\
& \tilde{J}_{\perp ,2} = j_{\perp,2} + i \xi
j_{\perp,1} , & \nonumber \\
&  \tilde{J}_\| = j_\| , &
\end{eqnarray}
with principal variances
\begin{eqnarray}
& ( \Delta \tilde{J}_{\perp ,1} )^2 = 0 , &
\nonumber \\
& ( \Delta \tilde{J}_{\perp ,2} )^2 \simeq
( \Delta \tilde{J}_\| )^2 \simeq \frac{1}{2}j(j+1). &
\end{eqnarray}
The vanishing of $\Delta \tilde{J}_1 $ is equivalent
to the eigenvalue equation (\ref{ee}).

\subsection{Schr\"{o}dinger cat states}

In this context a suitable example of Schr\"{o}dinger
cat states are the coherent superposition of two
opposite SU(2) coherent states. In the basis of
simultaneous eigenvectors of $j_0$ and a properly
chosen $j_3$ we have
\begin{equation}
\label{scs}
| \psi \rangle = \frac{1}{\sqrt{2}} \left (
|j,j \rangle + |j,-j \rangle \right ) ,
\qquad
j_0 | \psi \rangle = j | \psi \rangle  ,
\end{equation}
which for large $j$ are also known as maximally
entangled states, or NOON states, because of their
form in the number basis of two-mode bosonic
realizations, being also of much interest in
metrological applications \cite{noon}.

For $j =1/2$ these are SU(2) coherent states while
for $j \geq 1$ we have $\langle \bm{j} \rangle =
\bm{0}$ and $M$ and $\tilde{M}$ coincide
\begin{equation}
M = \tilde{M} = \pmatrix{ j/2 + \delta_{j,1}/2 & 0
& 0 \cr 0 & j/2 - \delta_{j,1}/2 & 0 \cr 0 &  0 &
j^2 } .
\end{equation}

It can be seen that $|\psi \rangle$ is a minimum
uncertainty state for the product and sum of
variances in Eqs. (\ref{prim2}) and (\ref{fap})
for $u,v=1,2$. On the other hand, the sum of three
variances takes the maximum value possible in
Eq. (\ref{Ds}). Moreover, the scaling of the
maximum principal variance as $(\Delta J_1 )^2
\propto j^2$ confirms the metrological usefulness
of these states \cite{noon}.

\subsection{States $|j,0 \rangle + |j,1 \rangle $ }

Finally, let us consider the following states expressed
in the basis of eigenvectors of $j_0$ and $j_3$ as
\begin{equation}
| \psi \rangle = \frac{1}{\sqrt{2}} \left (
|j,0 \rangle + |j,1 \rangle  \right ) ,
\end{equation}
with applications in quantum metrology \cite{HL}.
In our context these states provide an example
where the longitudinal component is not principal.
In this case
\begin{equation}
\langle \bm{j} \rangle = \frac{1}{2}
\pmatrix{\sqrt{j(j+1)}, & 0 , & 1},
\end{equation}
so that the longitudinal component is given by
\begin{equation}
j_\| = \sin \theta j_1 + \cos \theta j_3 ,
\qquad
\tan \theta = \sqrt{j(j+1)} .
\end{equation}
On the other hand, $M$ is diagonal in the $\bm{j}$
basis
\begin{equation}
M = \frac{1}{4} \pmatrix{ j(j+1)-1 & 0 & 0 \cr
0 & 2j(j+1)-1 & 0 \cr 0 & 0 & 1} ,
\end{equation}
and
\begin{equation}
\tilde{M} = \frac{1}{4}
\pmatrix{ j(j+1)-1 & i & 0 \cr
-i & 2j(j+1)-1 & i \sqrt{j(j+1)} \cr
0 & -i \sqrt{j(j+1)}& 1} ,
\end{equation}
so that no principal component coincides with $j_\|$.

\section{Conclusions}

In this work we have elaborated the assessment
of angular-momentum fluctuations via principal
variances derived from the diagonalization of
the covariance matrix for the problem. We have
considered two forms for the covariance matrix,
real symmetric and complex Hermitian. We have
related the principal variances with meaningful
SU(2) invariant uncertainty relations.

In particular we have derived nontrivial
uncertainty relations for states with vanishing
mean values of all angular-momentum components,
for which all previously introduced variance
products are trivially bounded by zero. We have found
that the corresponding minimum uncertainty states
are the maximally entangled states (NOON states
or Schr\"{o}dinger cat states). Moreover, we have
demonstrated that all pure states with vanishing
mean angular momentum are optimum for metrological
applications since they can reach the Heisenberg
limit.

\section*{Acknowledgment}

A. L. acknowledges the support from project
PR1-A/07-15378 of the Universidad Complutense.

\end{document}